\documentclass[english,final, journal]{IEEEtran}
\usepackage[T1]{fontenc}
\usepackage[latin9]{inputenc}
\usepackage{babel}
\usepackage{float}
\usepackage{amsmath}
\usepackage{graphicx}
\usepackage[unicode=true,
 bookmarks=true,bookmarksnumbered=true,bookmarksopen=true,bookmarksopenlevel=1,
 breaklinks=false,pdfborder={0 0 0},pdfborderstyle={},backref=false,colorlinks=false]
 {hyperref}
\hypersetup{pdftitle={Your Title},
 pdfauthor={Your Name},
 pdfpagelayout=OneColumn, pdfnewwindow=true, pdfstartview=XYZ, plainpages=false}
\usepackage{breakurl}

\makeatletter

\floatstyle{ruled}
\newfloat{algorithm}{tbp}{loa}
\providecommand{\algorithmname}{Algorithm}
\floatname{algorithm}{\protect\algorithmname}

 \let\oldforeign@language\foreign@language
 \DeclareRobustCommand{\foreign@language}[1]{%
   \lowercase{\oldforeign@language{#1}}}

\usepackage[caption=false,font=footnotesize]{subfig}
\usepackage{algorithmic}
\usepackage{balance}

\makeatother

\begin{document}

\title{Joint Channel Estimation and Nonlinear Distortion Compensation in
OFDM Receivers}

\author{Sergey V. Zhidkov,~\IEEEmembership{Member,~IEEE}\thanks{The author is with Cifrasoft Ltd., Izhevsk, Russia, e-mail: \protect\href{mailto:sergey.zhidkov@cifrasoft.com}{sergey.zhidkov@cifrasoft.com}.}}

\markboth{Joint Channel Estimation and Nonlinear Distortion Compensation in
OFDM Receivers}{}
\maketitle
\begin{abstract}
Nonlinear distortion in power amplifiers (PA) can significantly degrade
performance of orthogonal frequency division multiplexed (OFDM) communication
systems. This paper presents a joint maximum-likelihood channel frequency
response and nonlinear PA model estimator for OFDM signals. Derivation
of the estimator is based on Taylor-series representation of power
amplifier nonlinearity and is suitable for wide range of memoryless
PA models. A sub-optimal decision-aided algorithm for adaptive compensation
of nonlinear distortion effects at the receiver-side is also presented.
It is shown that the proposed algorithms can be used in IEEE 802.11a/g/p/ac
compliant wireless LAN receivers without any modifications at the
transmitter side. The performance of the proposed algorithms is studied
by means of computer simulation.
\end{abstract}


\section{Introduction}

\IEEEPARstart{I}{n recent years}, orthogonal frequency-division multiplexing
(OFDM) has emerged as a preferred candidate for a wide variety of
wireless communication applications. OFDM has been used in wireless
local area networks \cite{key-43}, in vehicular communication systems
\cite{key-42}, in European digital terrestrial video broadcasting
\cite{key-41}, and is a major contender for 5th generation mobile
networks \cite{key-40}. 

OFDM has several advantages over single-carrier systems, including
spectral effectiveness, robustness to multi-path propagation and efficient
implementation based on fast Fourier transform (FFT). Despite several
advantages OFDM has one major drawback - high sensitivity to nonlinear
distortions caused by the use of power amplifiers (PAs) at the transmitter
\cite{key-33,key-34,key-35,key-58,key-51}. The nonlinearity of PA
causes interference both inside and outside the OFDM signal bandwidth.
The out-of-band interference affects adjacent frequency channels,
whereas the in-band interference results in degradation of system
bit error rate (BER). Often, out-of-band spectral regrowth is a more
serious problem, but in high throughput wireless applications, BER
degradation may become unacceptable even when out-of-band spectral
regrowth is tolerable.

There are a number of methods that can be implemented in OFDM transmitters
in order to reduce performance degradation caused by PA nonlinearity.
These methods include deliberate clipping schemes \cite{key-50},
reduced peak-to-average power ratio (PAPR) coding \cite{key-20},
and amplifier pre-distortion techniques \cite{key-21}. However, such
techniques may be hard to implement, for example, in low-cost mobile
terminals for vehicular communication systems or wireless local area
network applications due to power, complexity or cost constraints.
In such a case, the receiver-side nonlinearity compensation may be
an attractive alternative for uplink processing, where more computational
resources are available at the base station.

Recent studies \cite{key-25,key-12} show that the nonlinear amplification
of OFDM signals combined with maximum-likelihood decoding at the receiver
may deliver better BER performance than that of linear OFDM transmission.
Unfortunately, true maximum-likelihood decoding it too complex for
practical implementation. Recently, there have been several studies
devoted to the sub-optimal reconstruction of nonlinearly distorted
OFDM signals at the receiver side \cite{key-13,key-14,key-15,key-17,key-18}.
These techniques permit implementation of nonlinear OFDM decoders
with intermediate complexity. Nonetheless, most studies assume that
the receiver knows the transmitter nonlinear transfer function, and
that a perfect channel state information is available at the receiver.
These assumptions are somewhat artificial, therefore previous research
has tended to focus on compensation of special types of nonlinearity,
such as deliberate clipping for PAPR reduction \cite{key-14,key-15,key-17},
rather than on realistic distortions introduced in PA. To the best
of author knowledge the problem of joint channel response and PA model
estimation has not been addressed in the previous studies. In \cite{key-18},
the authors propose the receiver-side nonlinearity compensation with
adaptive PA model estimation, without assuming perfect knowledge of
channel state information at the receiver. However, the authors in
\cite{key-18}, rely on conventional channel estimation techniques
developed for linear multipath channels, and therefore either require
special low-PAPR training symbols to minimize nonlinear distortion
effect on channel estimation or incur performance penalty due to imperfect
channel estimation. 

In this paper, we propose an adaptive channel estimation and nonlinear
distortion compensation algorithm for OFDM receivers. The proposed
algorithm does not assume a perfect channel state information and
prior knowledge of PA nonlinear transfer function at the receiver,
and can be used to jointly estimate and compensate the channel response
and PA nonlinearity using regular OFDM signal structure with block
type pilot arrangements, and also mitigate the nonlinear distortion
effects in decision-directed mode. The major difference between our
approach and previous studies is that we rely on frequency-domain
representation of PA nonlinearity. It simplifies derivation of joint
maximum likelihood channel and PA model estimator and permits nonlinear
distortion compensation solely in a frequency domain without costly
conversions from frequency-domain to time-domain representation on
every algorithm iteration. 

The paper is organized as follows. In Section II, the model of OFDM
system with nonlinear PA is defined. In Section III, a joint maximum
likelihood channel and PA model estimator and a sub-optimal iterative
decision-directed algorithm for detection of nonlinearly amplified
OFDM signals are presented and performance of the proposed algorithms
is studied by means of simulation in Section IV. Finally, Section
V draws conclusions. 

\section{System Model}

\begin{figure}[tbh]
\includegraphics[scale=1.1]{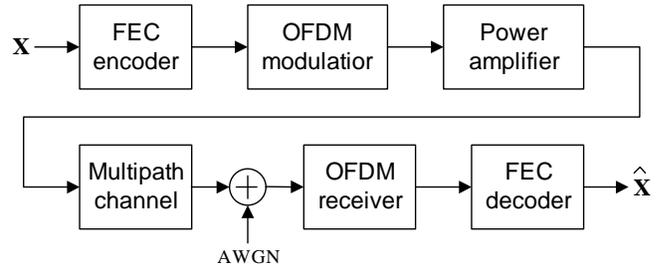}\caption{\label{fig:System-model}System model}
\end{figure}

Let us first introduce the OFDM transmission system shown in fig.
\ref{fig:System-model}. In the OFDM transmitter, information bits
are mapped into baseband symbols $\{S_{k}\}$ using $m$-ary phase-shift-keying
(PSK) or quadrature-amplitude-modulation (QAM) format. During active
symbol interval a block of $N$ complex baseband symbols ${\bf S}=[S_{0},S_{1},...,S_{N-1}]$
(possibly encoded by forward error correcting code) is transformed
by means of inverse discrete Fourier transform (IDFT) and digital-to-analog
conversion to the baseband OFDM signal as 

\begin{equation}
z(t)=\sum\limits _{k=0}^{N-1}S_{k}e^{j2\pi k\Delta ft},\quad0<t<T_{s},\label{eq:Baseband_analog}
\end{equation}

where $N$ is the number of sub-carriers, $\Delta f$ is the separation
between adjacent sub-carriers, and $T_{s}$ is the active symbol interval.
In practical OFDM systems, a cyclic prefix is usually added to every
symbol $z(t)$. The cyclic prefix is a periodic extension of the symbol
$z(t)$, which is primarily used to simplify equalizer design at the
receiver side. We also assume that an OFDM signal contains a block
of pilot symbols or set of pilot subcarriers to facilitate channel
estimation and carrier and timing recovery at the receiver. 

Two types of PA are mostly used in modern communication systems: travelling
waves tube amplifiers (TWTA) and solid-state power amplifiers (SSPA).
TWTA are used in high power satellite links while SSPA are used in
many other applications because of its small size. In this paper,
we mainly focus our attention on transmitters with memory-less nonlinearity.
Under such assumption the baseband signal distorted in a nonlinear
PA can be expressed as 
\begin{equation}
y\left(t\right)=F_{A}\left[\left|z\left(t\right)\right|\right]e^{j\left(\arg\left[z\left(t\right)\right]+F_{P}\left[\left|z\left(t\right)\right|\right]\right)},
\end{equation}
where $F_{A}[x]$ and $F_{P}[x]$ are the AM/AM and AM/PM functions
(AM/AM nonlinearity causes amplitude distortions which depend on amplitude
of the signal, while AM/PM nonlinearity causes phase distortions which
depend on amplitude of the signal). AM/AM and AM/PM nonlinearity for
SSPA is well represented by Rapp's model \cite{key-19}: 
\begin{equation}
F_{A}[\rho]=\rho\left[1+\left(\frac{\rho}{A_{sat}}\right)^{v}\right]^{-1\left/v\right.}\begin{array}{cc}
, & F_{P}[\rho]=0,\end{array}\label{eq:Rapp model}
\end{equation}
where $A_{sat}$ is the output saturation voltage and $v$ is the
smoothness factor. A typical value for $v$ is 2 - 4. 

Saleh \cite{key-20} and Ghorbani \cite{key-21} models are the two
alternative representations of PA nonlinear transfer function suitable
for description of TWTA and SSPA, respectively. 

These models have gained a lot of popularity because they represent
the PA nonlinearity by simple analytical expressions. Unfortunately,
these traditional models can only be applied to a narrow class of
PA, characterized by a regular shape of AM/AM and AM/PM functions.
Instead, for more general solution, we rely on memoryless polynomial
model (Taylor series expansion) to represent arbitrary PA nonlinearity
\cite{key-30,key-31}: 
\begin{equation}
y\left(t\right)=\sum\limits _{p=3,5,...}^{P}\beta_{p}\left[z\left(t\right)\right]^{\left(p+1\right)\left/2\right.}\left[z^{*}\left(t\right)\right]^{\left(p-1\right)\left/2\right.},\label{eq:Compex_powerseries}
\end{equation}
where $P$ is the highest order of nonlinearity, and $\{\beta_{p}\}$
is the baseband power-series coefficient.

It is shown \cite{key-32} that the complex power-series (\ref{eq:Compex_powerseries})
represents a general nonlinear transfer function and its odd-order
coefficients can be extracted from AM/AM and AM/PM measurements. It
is important to note that only the odd-order terms produce in-band
distortions. The even-order distortions are filtered out in zonal
filter and do not influence system bit or packet error rate.

\section{Proposed Algorithms}

\subsection{Effect of nonlinear PA on transmitted OFDM symbols}

Let us consider the baseband OFDM signal at the output of nonlinear
PA modeled by (\ref{eq:Compex_powerseries}). Substituting (\ref{eq:Baseband_analog})
into (\ref{eq:Compex_powerseries}) we immediately obtain

\begin{align}
y(t) & =\beta_{1}\sum\limits _{k=0}^{N-1}S_{k}e^{j2\pi k\Delta ft}\nonumber \\
 & +\beta_{3}\sum\limits _{n_{1}=0}^{N-1}\sum\limits _{n_{2}=0}^{N-1}\sum\limits _{n_{3}=0}^{N-1}S_{n_{1}}S_{n_{2}}S_{n_{3}}^{*}e^{j2\pi(n_{1}+n_{2}-n_{3})\Delta ft}\nonumber \\
 & +...\label{eq:Analog BB model}
\end{align}

From (\ref{eq:Analog BB model}) one can easily conclude that the
OFDM transmitter with memoryless PA (\ref{eq:Compex_powerseries})
is equivalent to a linear OFDM transmitter that emmits modified baseband
symbols $S'_{k}$, $k=0,1,...,N-1$:

\begin{align}
S'_{k} & =\beta_{1}S_{k}+\beta_{3}\times\sum\limits _{n_{1}+n_{2}-n_{3}=k}S_{n_{1}}S_{n_{2}}S_{n_{3}}^{*}\nonumber \\
 & +\beta_{5}\times\sum\limits _{{\scriptstyle {n_{1}+n_{2}+n_{3}\atop -n_{4}-n_{5}=k}}}S_{n_{1}}S_{n_{2}}S_{n_{3}}S_{n_{4}}^{*}S_{n_{5}}^{*}+...\label{eq:Digital BB label}
\end{align}

Note that $S'_{k}\ne0$ for $k<0$ and $k>N-1$, which results in
out-of-band emission. However, we do not consider the out-of-band
components and assume that these are filtered out in transmitter and/or
receiver filters.

It can be noted that the OFDM sub-carriers after nonlinear transformation
are no longer orthogonal; therefore the optimal maximum-likelihood
(ML) receiver requires a joint detection of transmitted vector ${\bf S}=[S_{0},S_{1},...,S_{N-1}]$
(see \cite{key-25,key-12} for detailed discussion). Unfortunately,
the optimal ML solution has little practical value. It cannot be used
in OFDM systems with large and intermediate number of sub-carriers
due to extremely high complexity.

Providing that $N$ is sufficiently large, a frequency-domain OFDM
symbol distorted in nonlinear PA (\ref{eq:Digital BB label}) can
be represented on the basis of extended Bussgang theorem \cite{key-33,key-34,key-39}
as

\begin{equation}
{\bf S'}=\alpha{\bf S}+{\bf d},\label{eq:Distortion simple}
\end{equation}
where ${\bf S'}=\left[{S'_{0},S'_{1},...,S'_{N-1}}\right]$, $\alpha$
is the complex attenuation factor, and ${\bf d}=\left[{d_{0},d_{1},...,d_{N-1}}\right]$
is the uncorrelated nonlinear distortion term.

After transformation in multipath channel and conventional DFT-based
demodulation the received signal vector ${\bf R}=\left[{R_{0},R_{1},...,R_{N-1}}\right]$
can be expressed as 

\begin{equation}
{\bf R}=\mathbf{\mathbf{\Theta}}{\bf S'}+{\bf w}=\alpha{\bf \mathbf{\Theta}S}+\mathbf{\Theta}{\bf d}+{\bf w},
\end{equation}
where ${\bf w}=[w_{0},w_{1},...,w_{N-1}]$ is the complex white Gaussian
noise vector with i.i.d. components having zero-mean and variance
$\sigma_{w}^{2}$, and $\mathbf{\Theta}=diag\left(\left[H_{0},H_{1},...,H_{N-1}\right]\right)$
is the diagonal matrix containing the frequency domain channel coefficients
(i.e. discrete Fourier transform of the channel impulse response).

Consider the additive distortion term ${\bf d}$. From (\ref{eq:Distortion simple})
it is straightforward to express ${\bf d}$ as

\begin{equation}
{\bf d}={\bf S'}-\alpha{\bf S}.
\end{equation}

Now, using (\ref{eq:Digital BB label}) we can represent ${\bf S'}$
in a compact form

\begin{equation}
{\bf S'}=\beta_{1}{\bf S}+\sum\limits _{p=3,5,...}^{P}{\beta_{p}{\bf d}^{\left(p\right)}},
\end{equation}
where elements of ${\bf d}^{\left(p\right)}=\left[d_{0}^{\left(p\right)},d_{1}^{\left(p\right)},...,d_{N-1}^{\left(p\right)}\right]$
can be computed by means of the discrete convolution:

\begin{equation}
\begin{array}{c}
\{d_{k}^{(3)}\}=\{S_{k}\}*\{S_{k}\}*\{S_{N-k}^{*}\},\\
\{d_{k}^{(5)}\}=\{S_{k}\}*\{S_{k}\}*\{S_{k}\}*\{S_{N-k}^{*}\}*\{S_{N-k}^{*}\},\\
...\\
\{d_{k}^{(P)}\}=\underbrace{\{S_{k}\}*\{S_{k}\}*...*\{S_{N-k}^{*}\}}_{P-{\rm {sequences}}},
\end{array}\label{eq:Convolutions}
\end{equation}
and $\{S_{N-k}^{*}\}$ represents the complex conjugated and reversely
ordered sequence of $\{S_{k}\}$. Note that ${\bf d}^{\left(p\right)}$
can efficiently be computed by zero-padding sequences $\{S_{k}\}$
and $\{S_{N-k}^{*}\}$, taking IDFT, multiplying appropriate coefficients
of IDFT, taking DFT and removing out-of-band components \cite{key-36}.
It is also worth noting that computational complexity of convolutions
(\ref{eq:Convolutions}) can further be reduced by taking into account
a finite resolution of $\{S_{k}\}$ (typically 1-4 bit).

After simple manipulations, it is straightforward to obtain

\begin{equation}
{\bf d}=\alpha\left[\left({c_{1}-1}\right){\bf S}+\sum\limits _{p=3,5,...}^{P}{c_{p}{\bf d}^{\left(p\right)}}\right],\label{eq:Distortion via c}
\end{equation}
where $c_{p}=\beta_{p}\alpha^{-1}$ is the $p$-th normalized coefficient
of PA transfer function. As one can see from (\ref{eq:Distortion via c})
the distortion term is a function of transmitted symbol ${\bf S}=[S_{0},S_{1},...,S_{N-1}]$,
and $(P+1)/2$ unknown coefficients $c_{1},c_{3},...,c_{P}$. Conversely,
the complex PA gain $\alpha$ can be expressed using baseband power-series
coefficients $\{\beta_{p}\}$ as

\begin{equation}
\alpha=\beta_{1}+\sum\limits _{p=3,5,...}^{P}{\beta_{p}T^{\left(p\right)}},\label{eq:alpha_original}
\end{equation}
where term $T^{(p)}$ represents an average energy of all $p$-th
order distortion terms that produce scaled replica of information-bearing
signal ${\bf S}$. Generally, value of $T^{(p)}$ depends on the number
of subcarriers and the modulation scheme, and can be evaluated either
analytically, or numerically. For constant amplitude modulation schemes
(such as $m$-ary PSK), $T^{(p)}$ coincides with the number of all
$p$-th order distortion terms that produce scaled replica of ${\bf S}$
(see \cite{key-58}, for $p=3,5,7,9$ and $p\to\infty$). Derivation
of $T^{(p)}$ for $p=3,5$ and non-constant amplitude modulation (e.g.
$m$-ary QAM) is given in Appendix A. 

Dividing both sides of (\ref{eq:alpha_original}) by $\alpha$, we
can express coefficient $c_{1}$ as

\begin{equation}
c_{1}=1-\sum\limits _{p=3,5,...}^{P}{c_{p}T^{\left(p\right)}},\label{eq:c1 coeff}
\end{equation}
and by substituting (\ref{eq:c1 coeff}) in (\ref{eq:Distortion via c})
the distortion term can be expressed as

\begin{equation}
{\bf d}=\alpha\sum\limits _{p=3,5,...}^{P}{c_{p}\left({\bf d}^{(p)}-T^{(p)}{\bf S}\right)}.\label{eq:Distortion sum}
\end{equation}

Finally, the frequency-domain received signal can be reformulated
using (\ref{eq:Distortion sum}) as:

\begin{equation}
{\bf R}=\mathbf{\mathbf{\Theta}_{\alpha}}\left({\bf S}+\sum\limits _{p=3,5,...}^{P}{c_{p}\left({\bf d}^{(p)}-T^{(p)}{\bf S}\right)}\right)+{\bf w},\label{eq:Rec signal}
\end{equation}
where $\mathbf{\Theta_{\alpha}}=diag\left(\left[\alpha H_{0},\alpha H_{1},...,\alpha H_{N-1}\right]\right)$.

\subsection{Joint maximum-likelihood estimation of channel response and normalized
power-series coefficients}

To simplify further derivations we assume that the time-domain channel
impulse response (CIR) is not limited by cyclic prefix duration and,
therefore, we do not take into account correlation of elements in
vector ${\bf H}=\left[H_{0},H_{1},...,H_{N-1}\right]$. 

We aim to find a joint maximum likelihood estimator of vectors ${\bf H'}=\left[\alpha H_{0},\alpha H_{1},...,\alpha H_{N-1}\right]$
and ${\bf c}=[c_{3},c_{5},...,c_{P}]$, given a group of $M$ received
OFDM symbols ${\bf R}^{(m)}=[R_{0}^{(m)},R_{1}^{(m)},...,R_{N-1}^{(m)}]$,
$m=1,2...M$. Since the noise term ${\bf w}$ in (\ref{eq:Rec signal})
is i.i.d Gaussian distributed, the maximum-likelihood estimator is
equivalent to least-squares estimator that minimizes the error function
\cite{key-55}:

\begin{equation}
J=\sum\limits _{m=1}^{M}\left\Vert {\bf R}^{(m)}-\mathbf{\mathbf{\Theta_{\alpha}}}\left({\bf S}^{(m)}+{\bf U}^{(m)}{\bf c}\right)\right\Vert ^{2}\label{eq:J}
\end{equation}
where

\begin{equation}
{\bf U}^{(m)}={\bf d}^{(p,m)}-T^{(p)}{\bf S}^{(m)}
\end{equation}

Condition (\ref{eq:J}) leads to the system of $M\times N$ nonlinear
equations for $N+(P-1)/2$ unknown parameters:

\begin{align}
R_{k}^{(m)}-H'_{k}\left(S_{k}^{(m)}+\sum\limits _{p=3,5,...}^{P}c_{p}U^{(p,m)}\right),\label{eq:Normal eq system}\\
k=0,1,...,N-1\nonumber \\
m=1,2,...,M\nonumber 
\end{align}
where $U^{(p,m)}=d_{k}^{(p,m)}-T^{(p)}S_{k}^{(m)}$. Note that at
least two \emph{different} OFDM symbols are required to jointly estimate
channel frequency response and normalized power-series coefficients
since the total number of equations in (\ref{eq:Normal eq system})
should be no less than $N+(P-1)/2$.

Unfortunately, the solution of (\ref{eq:Normal eq system}) cannot
be expressed in a simple closed-form. Nonetheless, sub-optimal iterative
methods can be used to solve (\ref{eq:Normal eq system}). Here, we
propose one such method. 

Firstly, we note that the channel response vector ${\bf H'}$ that
minimizes $J$ for given ${\bf c}$ can be found as:

\begin{align}
\hat{H}'_{k}=\frac{1}{M}\sum\limits _{m=1}^{M}R_{k}^{(m)}\left(S_{k}^{(m)}+\sum\limits _{p=3,5,...}^{P}c_{p}U^{(p,m)}\right)^{-1},\label{eq:Solution for H}\\
k=0,1,...,N-1\nonumber 
\end{align}

One can easily observe that equation (\ref{eq:Solution for H}) is
equivalent to a conventional least-squares channel estimation, albeit
using the modified reference symbols $S_{k}^{(m)}+\sum\limits _{p=3,5,...}^{P}c_{p}\left(d_{k}^{(p,m)}-T^{(p)}S_{k}^{(m)}\right)$
and averaged over $M$ OFDM training symbols. On the other hand, the
power-series coefficient vector ${\bf c}$ that minimizes $J$ for
given ${\bf H'}$ can be found using linear least-squares solution:

\begin{equation}
{\bf \hat{c}}=\left({\bf U}^{H}{\bf \mathbf{\boldsymbol{\Lambda}}}^{H}{\bf \boldsymbol{\Lambda}U}\right)^{-1}{\bf U}^{H}{\bf \boldsymbol{\Lambda}}^{H}\left({\bf R}-{\bf \boldsymbol{\Lambda}S}\right)\label{eq:Solution for c}
\end{equation}
where ${\bf U}=\left[\mathbf{U}^{(1)},\mathbf{U}^{(2)},\mathbf{...,U}^{(M)}\right]^{T}$,
${\bf R}=\left[\mathbf{R}^{(1)},\mathbf{R}^{(2)},\mathbf{...,R}^{(M)}\right]^{T}$
, ${\bf S}=\left[\mathbf{S}^{(1)},\mathbf{S}^{(2)},\mathbf{...,S}^{(M)}\right]^{T}$,
and ${\bf \boldsymbol{\Lambda}}=diag[H'_{0},H'_{1},...,H'_{N-1},H'_{0},...,H'_{N-2},H'_{N-1}]$.

Then the solution of (\ref{eq:Normal eq system}) can be approximated
using the following iterative algorithm:

\begin{algorithm}[H]
\begin{enumerate}
\item Set the initial estimate ${\bf \hat{c}}=0$
\item Calculate the channel frequency response estimate ${\bf \hat{H}'}$
in accordance with (\ref{eq:Solution for H}) using current estimate
of ${\bf \hat{c}}$
\item Calculate the estimate of power-series coefficients ${\bf c}$ in
accordance with (\ref{eq:Solution for c}) using current estimate
of ${\bf \hat{H}'}$ 
\item Repeat steps 2) and 3) until the error function $J$ is no longer
decreasing or the maximum number of iterations is reached
\end{enumerate}
\caption{Joint estimation of channel response and PA model}

\end{algorithm}

\begin{figure}[t]
\includegraphics[scale=0.57]{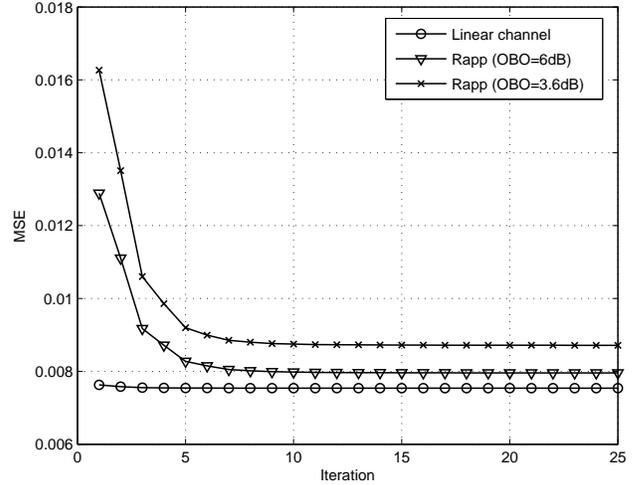}\caption{\label{fig:Convergence}Convergence of joint ML channel and PA model
estimator (Rapp PA model, \emph{v}=2,\emph{ N}=120, \emph{M}=2, \emph{SNR}=20
dB)}
\end{figure}

Although there is no guarantee that the proposed iterative procedure
will always converge to the global minimum, our numerical simulations
indicate that in most practical scenarios (i.e. typical noise levels,
multipath channel models, and amplifier nonlinearity) the proposed
iterative algorithm quickly converges to a steady-state solution that
minimizes (\ref{eq:J}). Learning curves obtained via simulation for
a system with two training symbols at different values of output back-off
(OBO) are illustrated in figure \ref{fig:Convergence}. In most cases,
the proposed procedure converges to a steady-state value of $J$ within
4-5 full iterations. 

\begin{figure}[t]
\includegraphics[scale=0.57]{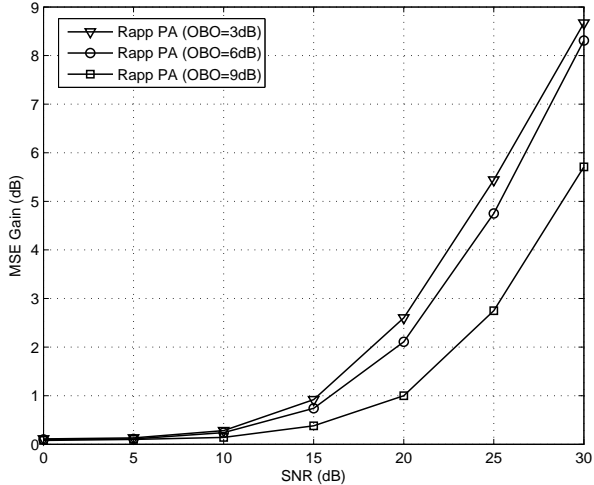}\caption{\label{fig:MSE-gain}MSE gain of the proposed joint channel and PA
model estimator over a conventional least-squares channel estimator
(Rapp PA model, \emph{v}=2,\emph{ N}=120, \emph{M}=2)}

\end{figure}

The MSE gain that can be achieved by using the proposed joint channel
and PA model estimator over a conventional least-squares channel estimator
in nonlinear channel (Rapp PA model) is illustrated in fig. \ref{fig:MSE-gain}.
Not surprisingly, the improvement in terms of MSE is higher in low
OBO and high SNR region where the MSE of the conventional least-squares
channel estimator is dominated by nonlinear distortion noise. 

It should be noted that the proposed joint maximum-likelihood estimation
of channel frequency response and normalized power-series coefficients
has moderate complexity. If the pilot symbols $\mathbf{S}$ are known
\emph{a priori}, the distortion vectors $\mathbf{U}^{(1)}$,$\mathbf{U}^{(2)}$,...,$\mathbf{U}^{(M)}$,
can be pre-computed. The channel frequency response calculation step
(\ref{eq:Solution for H}) requires only $N$ complex divisions, and
since $P$ is usually relatively small, the calculation of (\ref{eq:Solution for c})
is also fairly simple. For example, for $P=5$, the term ${\bf U}^{H}{\bf \boldsymbol{\Lambda}}^{H}{\bf \boldsymbol{\Lambda}U}$
in (\ref{eq:Solution for c}) is a 2x2 matrix. Furthermore, since
we estimate both channel frequency response and power amplifier coefficients
in frequency domain there are no costly DFT/IDFT operations at every
iteration of the algorithm. 

\subsection{Decision-aided nonlinear distortion compensation}

After applying a conventional zero-forcing equalization to the received
signal (\ref{eq:Rec signal}) the equalizer output (without Gaussian
noise term for simplicity) can be expressed as

\begin{equation}
R_{k}^{(eq)}=\frac{R_{k}}{H'_{k}}=S_{k}+\sum\limits _{p=3,5,...}^{P}c_{p}\left(d_{k}^{(p)}-T^{(p)}S_{k}\right)\label{eq:Equalized signal}
\end{equation}

The distortion term in (\ref{eq:Equalized signal}) can be canceled
using an iterative decision-aided approach \cite{key-13}. At the
first step, the equalized signal (\ref{eq:Equalized signal}) is utilized
to obtain tentative decisions ${\bf \hat{S}}$, e.g. using slicer
or a feedback from FEC decoder. Then, we can reconstruct distortion
term $\sum\limits _{p=3,5,...}^{P}\hat{c}_{p}\left(\hat{d}_{k}^{(p)}-T^{(p)}\hat{S}_{k}\right)$
and subtract it from the output of zero-forcing equalizer (\ref{eq:Equalized signal}).
Since the tentative decisions may contain errors the reconstructed
distortion term might be inaccurate. However, if the number of errors
in ${\bf \hat{S}}$ is relatively small the compensated signal will
be less noisy than the original equalizer output and the decisions
obtained at the second iteration will contain fewer errors. The proposed
decision-aided nonlinearity compensation algorithm can be summarized
as follows:

\begin{algorithm}[H]
\begin{enumerate}
\item Use equalized signal (\ref{eq:Equalized signal}) to obtain tentative
decisions ${\bf \hat{S}}$ (via slicer of FEC decoder)
\item Calculate ${\bf \hat{d}}^{\left(p\right)}=\left[\hat{d}_{0}^{\left(p\right)},\hat{d}_{1}^{\left(p\right)},...,\hat{d}_{N-1}^{\left(p\right)}\right],\;p=3,5,...,P$. 
\item Compensate the nonlinear distortion term $R_{k}^{(comp)}=R_{k}^{(eq)}-\sum\limits _{p=3,5,...}^{P}\hat{c}_{p}\left(\hat{d}_{k}^{(p)}-T^{(p)}\hat{S}_{k}\right),\;k=0,1,...,N-1$
\item Use distortion compensated signal $R_{k}^{(comp)}$to obtain updated
decisions ${\bf \hat{S}}$ 
\item Repeat steps 2)-4) until ${\bf \hat{S}}$ is no longer differs from
${\bf \hat{S}}$ obtained at previous iteration or until the maximum
number of iterations is reached.
\end{enumerate}
\caption{Iterative decision-aided nonlinearity mitigation}

\end{algorithm}

In addition to distortion compensation, the decision-aided approach
can be also used to update estimate of ${\bf \hat{c}}$. In particular,
it is possible to update non-linear model parameters at each iteration
of the algorithm using (\ref{eq:Solution for c}) with $M$=1 and
tentative decision vector ${\bf \hat{S}}$. However, due to possibility
of errors in tentative decisions ${\bf \hat{S}}$ it is preferable
to average these estimates over several OFDM symbols to reduce the
effect of decision errors in a single OFDM symbol. One simple way
to achieve this it is to employ the first order recursive filter

\begin{equation}
{\bf \hat{c}}_{t}^{(avg)}=\gamma\cdot{\bf \hat{c}}_{t}+(1-\gamma)\cdot{\bf \hat{c}}_{t-1}^{(avg)},\label{eq:smoothing c}
\end{equation}
where where $t$ is the OFDM symbol index, and $\gamma$ is the smoothing
factor. The optimal choice of $\gamma$ shall be discussed in the
next section.

\subsection{Limitations}

The proposed channel estimation and nonlinear distortion compensation
technique has a few limitations:
\begin{itemize}
\item Firstly, the Bussgang theorem is only applicable if the number of
OFDM sub-carriers $N\rightarrow\infty$. Yet, many state-of-the-art
wireless communication schemes employ OFDM modulation with relatively
small number of subcarriers (e.g. in IEEE802.11a/g/p, the number of
active subcarriers is $N$=52). From (\ref{eq:alpha_k}) one can easily
see that for non-constant amplitude modulation (such as $m$-ary QAM)
and finite number of OFDM subcarriers the scaling factor $\alpha$
in (\ref{eq:Distortion simple}) is not a constant value as suggested
by Bussgang theory, but varies from symbol-to-symbol and from subcarrier-to-subcarrier. 
\item Secondly, the polynomial model parameters $\mathbf{\hat{c}}$ are
fitted to the training data that may have limited dynamic range, which
in turn may result in model underfitting. In particular, many packet-based
OFDM systems rely on training symbols with reduced PAPR to decrease
channel estimation noise caused by nonlinearities in conventional
channel estimators. Such arrangement, although beneficial for conventional
channel estimation schemes, limits the efficiency of the proposed
joint channel and non-linear model estimation algorithm, because the
nonlinear PA model may be incorrectly reconstructed outside of the
dynamic range represented by training symbols. 
\item Thirdly, since the proposed nonlinear distortion compensation technique
relies on decision feedback mechanism, it is susceptible to error
propagation effects. In fact, if the first tentative decision vector
$\mathbf{\hat{S}}$ contains too many errors the reconstructed distortion
term may significantly differ from a true distortion term and the
SNR of the compensated signal $R_{k}^{(comp)}$ may become even worse
than the SNR at the output of the conventional zero-forcing equalizer
(\ref{eq:Equalized signal}). 
\end{itemize}
Several ad-hoc techniques can be employed to reduce the error propagation
effect. For example, one can utilize decision feedback from FEC decoder.
However, in such a case, caution should be used to avoid further deterioration
of performance due to error correlation. For example, the bit errors
produced by Viterbi decoding are correlated and tend to group together
in error bursts. Therefore, if the number of bits transmitted per
OFDM symbol is relatively small and there is no interleaving between
OFDM symbols the hard-decision feedback from Viterbi decoder may significantly
deteriorate performance of the proposed decision-aided nonlinear distortion
compensation algorithm. Another simple method to partially reduce
the error-propagation effect is to intentionally decrease the compensation
term at first iterations of the decision-aided algorithm, i.e. $R_{k}^{(comp)}=R_{k}^{(eq)}-\mu_{i}\sum\limits _{p=3,5,...}^{P}\hat{c}_{p}\left(\hat{d}_{k}^{(p)}-T^{(p)}\hat{S}_{k}\right),\;k=0,1,...,N-1$,
where $\mu_{i}\leq1$ is the damping factor at \emph{i}-th iteration,
such that $\mu_{i}\leq\mu_{i+1}$ . This technique has demonstrated
some performance gain in OFDM systems with small number of subcarriers
(see simulation section for details). 

\subsection{Application of the proposed technique to IEEE802.11ac compliant receivers}

The proposed joint channel and non-linear model estimation and distortion
compensation scheme can be used in standard-complaint IEEE802.11ac
receivers with minor modifications. We consider a system with single
transmit chain, because in practice it is very likely that the power
amplifier nonlinearity will occur in low-cost mobile terminals that
usually rely on a single transmitter due to power, size and cost constraints.
The proposed channel and nonlinear model estimation technique requires,
at least, two different OFDM training symbols for initial channel
estimation. To satisfy this condition in IEEE802.11a/g/p or in legacy
mode of IEEE802.11ac \cite{key-43}, one can use the L-LTF training
symbol and the demodulated and reconstructed L-SIG symbol. Similarly,
in very high throughput (VHT) mode, one can use the VHT-LTF and SIG-B
symbols for initial channel and nonlinear model estimation. L-SIG
and SIG-B symbols are not known \emph{a priori} at the receiver-side,
however, they employ very robust 1/2-rate convolutional coding and
BPSK modulation \cite{key-43}, and therefore, can be demodulated
and reconstructed without errors in all channel conditions that may
be deemed suitable for 64-QAM and 256-QAM modes. After the initial
channel and nonlinear model estimation the nonlinear model parameters
${\bf \hat{c}}$ can be constantly updated in decision-directed way
as described in the previous section. 

\section{Simulation Results and Discussion}

To evaluate performance of the proposed scheme we simulated IEEE802.11ac
compliant single-input single-output (SISO) transceiver in VHT mode
and in legacy mode (thus, the results are also applicable to IEEE802.11a/g/p).
Perfect carrier frequency offset and symbol timing synchronization
is assumed. We primarily focus on 64-QAM and 256-QAM modes, because
in these modes the power amplifier nonlinearity may cause significant
bit and packet error rate degradation even though the spectral regrowth
may be well within spectral mask requirements \cite[p.298]{key-43}.
In all simulations, we assume that the channel and PA model estimator
is initialized at every transmitted burst. PA nonlinearity was modeled
by Rapp model (\ref{eq:Rapp model}) with parameter $v=2$. For fair
comparison, the linear receiver also relies on two OFDM symbols L-LTF
(or VHT-LTF) and the reconstructed L-SIG (or SIG-B) for least-squares
channel estimation. We simulated the receiver performance in AWGN
and block fading multipath channels with delay profile models proposed
in \cite{key-59} for typical WLAN indoor environment. The packet
size in all simulation cases was set to 400 bytes. 

\begin{figure}[t]
\includegraphics[scale=0.57]{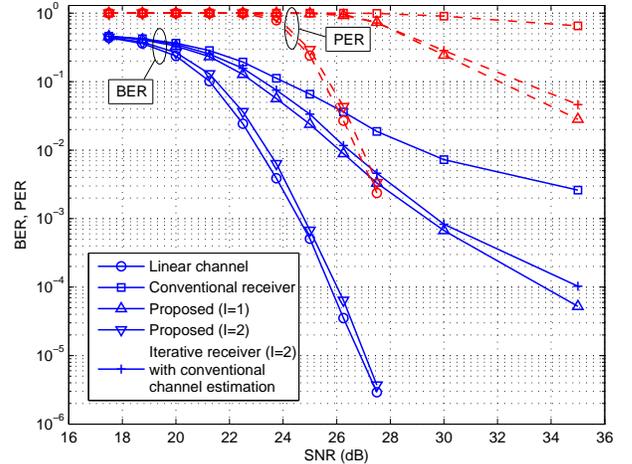}\caption{\label{fig:BER 256-QAM 160MHz AWGN}BER/PER vs SNR for IEEE802.11ac
compliant receiver (\emph{N}=484, 256-QAM, code rate 3/4) in AWGN
channel, Rapp PA (\emph{v}=2, OBO=9.7dB) }
\end{figure}

\begin{figure}[t]
\includegraphics[scale=0.57]{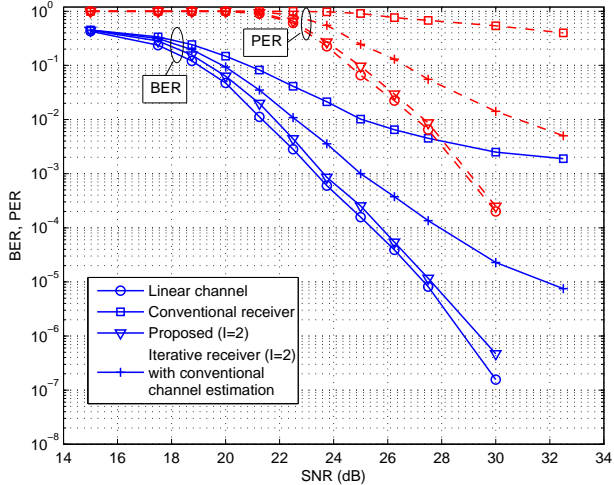}\caption{\label{fig:BER 64-QAM 160Mhz Ch B}BER/PER vs SNR for IEEE802.11ac
compliant receiver (\emph{N}=484, 64-QAM, code rate 3/4) in block
fading multipath channel (Model B \cite{key-59}), Rapp PA (\emph{v}=2,
OBO=8.6dB)}
\end{figure}

Simulation results for VHT mode (160MHz bandwidth and 484 active subcarriers)
are presented in figures \ref{fig:BER 256-QAM 160MHz AWGN} and \ref{fig:BER 64-QAM 160Mhz Ch B}.
As one can see the proposed algorithms allow almost perfect compensation
of mild nonlinear distortions in IEEE802.11ac VHT mode both in AWGN
and typical multipath channels. The proposed receiver performance
is just a fraction of dBs away from linear transmission after two
or three iterations. It should be noted that the proposed joint channel
and PA model estimation plays a key role in the overall performance
improvement. In particular, the decision-aided iterative nonlinearity
compensation algorithm combined with a conventional channel estimation
scheme demonstrates substantially worse performance than the proposed
receiver with joint channel and PA model estimation (fig. \ref{fig:BER 256-QAM 160MHz AWGN},
\ref{fig:BER 64-QAM 160Mhz Ch B}).

\begin{figure}[t]
\includegraphics[scale=0.57]{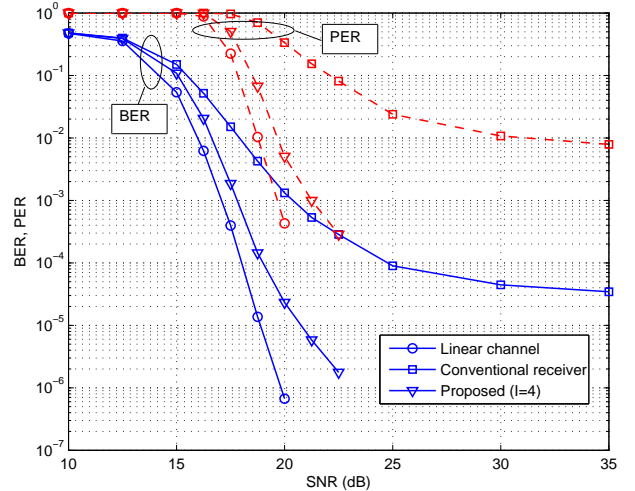}\caption{\label{fig:BER legacy AWGN}BER/PER vs SNR for IEEE802.11ac compliant
receiver in legacy mode (\emph{N}=52, 64-QAM, code rate 3/4), AWGN,
Rapp PA (\emph{v}=2, OBO=8.5dB)}
\end{figure}

\begin{figure}[t]
\includegraphics[scale=0.57]{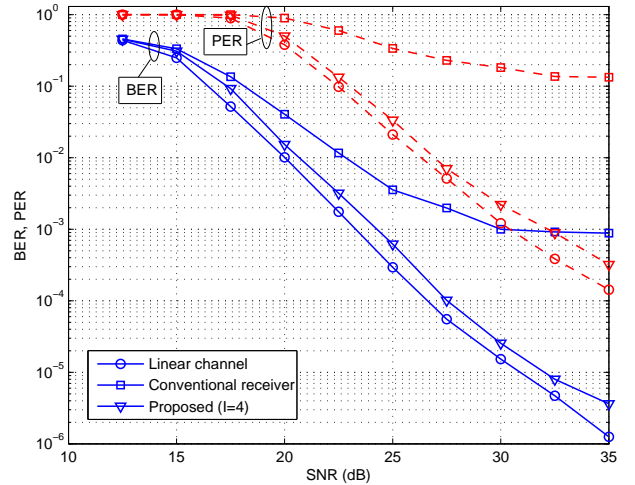}\caption{\label{fig:BER legacy Ch B}BER/PER vs SNR for IEEE802.11ac compliant
receiver in legacy mode (\emph{N}=52, 64-QAM, code rate 3/4) in block
fading multipath channel (Model A \cite{key-59}), Rapp PA (\emph{v}=2,
OBO=8.5dB)}
\end{figure}

\begin{figure}[t]
\includegraphics[scale=0.57]{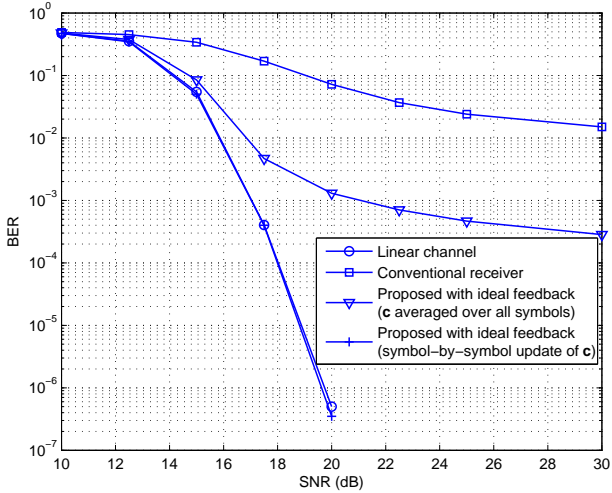}\caption{\label{fig:BER ideal feedback}BER vs SNR for IEEE802.11ac compliant
receiver in legacy mode (\emph{N}=52, 64-QAM, code rate 3/4) in AWGN,
Rapp PA (\emph{v}=2, OBO=6dB) with ideal decision feedback}
\end{figure}

Simulation results for legacy mode (20MHz bandwidth and 52 active
subcarriers) are presented in fig. \ref{fig:BER legacy AWGN} and
\ref{fig:BER legacy Ch B}. In legacy mode, the performance improvement
is also significant, although the overall performance is slightly
worse than that in VHT mode. In addition, to get the best results
it was necessary to increase the number of iterations and downscale
the compensation term at the first iteration ($\mu_{1}=0.75$) to
minimize the adverse effect of error propagation. The main reason
for slightly worse performance in legacy mode is the lower number
of active subcarriers per OFDM symbol. The effect of reducing the
number of active subcarriers is two-fold. Firstly, a fewer data bits
per OFDM symbol means that accidental tentative decision errors at
the first step of decision-aided compensation algorithm may have larger
impact on algorithm convergence. Secondly, as we discussed earlier,
in case of small number of OFDM subcarriers, the Bussgang theorem
is no longer valid, and as a result the scaling factor $\alpha$ in
(\ref{eq:Distortion simple}) is no longer constant, but varies from
OFDM symbol to OFDM symbol. It means that the optimal set of coefficients
${\bf \hat{c}}$ that minimizes distortion function (\ref{eq:J})
for training OFDM symbols may be highly sub-optimal for some OFDM
data symbols. This fact is illustrated in figure \ref{fig:BER ideal feedback},
where we plot simulation results for legacy mode (\emph{N}=52) using
ideal decision feedback instead of slicer decisions. One may expect
that the ideal decision feedback should always result in a perfect
compensation of nonlinear distortions, however, this is not the case.
If estimation of ${\bf \hat{c}}$ is averaged over all OFDM symbols
in a packet, the performance curves exhibit error-floor behavior at
relatively high BER/PER. The issue can be solved by re-estimating
${\bf \hat{c}}$ at every OFDM symbol independently, which is equivalent
to setting $\gamma=1$ in (\ref{eq:smoothing c}). However, this approach
demonstrates poor performance with non-ideal decision feedback, since
relying on incorrect decisions ${\bf \hat{S}}$ may significantly
affect estimation of parameter vector ${\bf \hat{c}}$ and may lead
to considerable performance degradation. Therefore, the optimal smoothing
factor $\gamma$ was found experimentally; in most cases, it was found
to be in the range of $\gamma=0.2\div0.3$. 

It should be noted that the proposed algorithm is based on memoryless
model of power amplifier. This approach is mainly justified by the
fact that the low-power amplifiers used in mobile terminals usually
exhibit very weak memory effects. However, the proposed technique
can be also applied to the compensation of nonlinear distortion effects
caused by PA with memory. In particular, a combination of memoryless
PA and frequency-selective fading channel can be viewed as a Hammerstein
nonlinear model, which is often used to model memory effects in PAs
and demonstrates good modeling behavior \cite{key-5}. Moreover, our
simulation results (not shown here for brevity) indicate that the
proposed joint channel estimation and distortion compensation technique
provides substantial performance gain for Wiener-type PA models.

\section{Conclusions}

Novel algorithms for joint channel response and PA model estimation
and decision-aided nonlinear distortion compensation for nonlinearly
distorted OFDM signals have been proposed. The proposed algorithms
are suitable for a typical OFDM communication system with block type
pilot arrangement. In particular, they can be used in standard-compliant
IEEE802.11ac receivers. The proposed receiver performance has been
evaluated in terms of bit and packet error rate through computer simulations,
demonstrating the significant benefits of the proposed algorithms
over conventional linear receivers, especially in high throughput
modes of operation.

\appendices{}

\section{Complex attenuation factor and uncorrelated nonlinear distortion
term ($p$=5)}

Consider $k$-th subcarrier of OFDM symbol transformed in nonlinear
PA modeled by fifth-order polynomial model. Its representation is
given by (\ref{eq:Digital BB label}). It can be observed that the
third- and fifth-order terms on the right-hand side of equation (\ref{eq:Digital BB label})
produce the scaled replica of transmitted symbol $S_{k}$ if one of
the following conditions is met:
\begin{enumerate}
\item For the third-order terms:\\
 $\ensuremath{A_{1}^{(3)}}:\ensuremath{\left([n_{1}=k]\&[n_{2}=n_{3}]\right)}$
or\\
 $\ensuremath{A_{2}^{(3)}}:\ensuremath{\left([n_{2}=k]\&[n_{1}=n_{3}]\right)}$
\item For the fifth-order terms: \\
$\ensuremath{A_{1}^{(5)}:\left([n_{1}=k]\&[n_{2}=n_{4}]\&[n_{3}=n_{5}]\right)}$
or \\
$\ensuremath{A_{2}^{(5)}}:\ensuremath{\left([n_{1}=k]\&[n_{2}=n_{5}]\&[n_{3}=n_{4}]\right)}$
or\\
$\ensuremath{A_{3}^{(5)}}:\ensuremath{\left([n_{2}=k]\&[n_{1}=n_{4}]\&[n_{3}=n_{5}]\right)}$
or\\
$\ensuremath{A_{4}^{(5)}}:\ensuremath{\left([n_{2}=k]\&[n_{1}=n_{5}]\&[n_{3}=n_{4}]\right)}$
or\\
$\ensuremath{A_{5}^{(5)}}:\ensuremath{\left([n_{3}=k]\&[n_{1}=n_{4}]\&[n_{2}=n_{5}]\right)}$
or\\
$\ensuremath{A_{6}^{(5)}}:\ensuremath{\left([n_{3}=k]\&[n_{1}=n_{5}]\&[n_{2}=n_{4}]\right)}$.
\end{enumerate}
The sum of all third-order terms for which at least one of conditions
$\ensuremath{A_{1}^{(3)}}$ or $\ensuremath{A_{2}^{(5)}}$ is met
can be calculated using the equality from combinatorial theory \cite{key-38}: 

\begin{equation}
\ensuremath{\left|A_{1}^{(3)}\cup A_{2}^{(3)}\right|=\left|A_{1}^{(3)}\right|+\left|A_{2}^{(3)}\right|-\left|A_{1}^{(3)}\cap A_{2}^{(3)}\right|}
\end{equation}

where $\left|\cdot\right|$ denotes the sum of all terms for which
given condition is met. If for given set of indexes $k$, $n_{1}$,
$n_{2}$, $n_{3}$ condition $\ensuremath{A_{1}^{(3)}}$ is met the
third-order term becomes

\begin{equation}
\ensuremath{\left|A_{1}^{(3)}\right|=\beta_{3}S_{k}\sum\limits _{t=0}^{N-1}\left|S_{t}\right|^{2}}
\end{equation}

The same result can be obtained for condition $\ensuremath{A_{2}^{(5)}}$.
It is easy to show that

\begin{equation}
\ensuremath{\left|A_{2}^{(3)}\right|=\left|A_{1}^{(3)}\right|}
\end{equation}

It can be noted that both conditions $\ensuremath{A_{1}^{(3)}}$ and
$\ensuremath{A_{2}^{(5)}}$ are met simultaneously if $\ensuremath{n_{1}=n_{2}=n_{3}=k}$.
For given $k$ there is only one third-order term, which satisfies
condition $\ensuremath{{n_{1}}={n_{2}}={n_{3}}=k}$. Thus, $\ensuremath{\left|A_{1}^{(3)}\cap A_{2}^{(3)}\right|}$
is determined by 

\begin{equation}
\ensuremath{\left|A_{1}^{(3)}\cap A_{2}^{(3)}\right|=\beta_{3}S_{k}\left|S_{k}\right|^{2}}
\end{equation}

Finally, the sum of all third-order terms, which produce scaled replica
of $S_{k}$, can be expressed as

\begin{equation}
\ensuremath{\left|A_{1}^{(3)}\cup A_{2}^{(3)}\right|=\beta_{3}S_{k}\left(2\sum\limits _{t=0}^{N-1}\left|S_{t}\right|^{2}-\left|S_{k}\right|^{2}\right)}\label{eq:third_order sum}
\end{equation}

Similarly, the sum of all fifth-order terms for which at least one
of conditions $\ensuremath{A_{i}^{(5)},}i=1,2,...,6$ is met can be
calculated using the following equality \cite{key-38}: 

\begin{equation}
\ensuremath{\begin{array}{c}
\left|A_{1}^{(5)}\cup A_{2}^{(5)}\cup A_{3}^{(5)}\cup A_{4}^{(5)}\cup A_{5}^{(5)}\cup A_{6}^{(5)}\right|\\
=\sum\limits _{i=1}^{6}\left|A_{i}^{(5)}\right|-\sum\limits _{i<j}\left|A_{i}^{(5)}\cap A_{j}^{(5)}\right|\\
+\sum\limits _{i<j<k}\left|A_{i}^{(5)}\cap A_{j}^{(5)}\cap A_{k}^{(5)}\right|\\
-\sum\limits _{i<j<k<t}\left|A_{i}^{(5)}\cap A_{j}^{(5)}\cap A_{k}^{(5)}\cap A_{t}^{(5)}\right|\\
+\sum\limits _{i<j<k<t<u}\left|A_{i}^{(5)}\cap A_{j}^{(5)}\cap A_{k}^{(5)}\cap A_{t}^{(5)}\cap A_{u}^{(5)}\right|\\
-\left|A_{1}^{(5)}\cap A_{2}^{(5)}\cap A_{3}^{(5)}\cap A_{4}^{(5)}\cap A_{5}^{(5)}\cap A_{6}^{(5)}\right|
\end{array}}
\end{equation}

Due to space limitation we present here only the final result of (31)
calculation. It can be shown that

\begin{align}
\ensuremath{\left|A_{1}^{(5)}\cup...\cup A_{6}^{(5)}\right|} & =\beta_{5}S_{k}\times6\sum\limits _{t=0}^{N-1}\sum\limits _{l=0}^{N-1}\left|S_{t}\right|^{2}\left|S_{l}\right|^{2}\label{eq:fifth_order sum}\\
 & -\beta_{5}S_{k}\times6\sum\limits _{t=0}^{N-1}\left|S_{t}\right|^{2}\left|S_{k}\right|^{2}\nonumber \\
 & -\beta_{5}S_{k}\times3\sum\limits _{t=0}^{N-1}\left|S_{t}\right|^{4}\nonumber \\
 & +\beta_{5}S_{k}\times4\left|S_{k}\right|^{4}\nonumber 
\end{align}

Now, making use of (\ref{eq:third_order sum}) and (\ref{eq:fifth_order sum})
we can rewrite expression (\ref{eq:Digital BB label}) as

\begin{equation}
\ensuremath{S'_{k}=\alpha_{k}S_{k}+d_{k}}
\end{equation}
where $d_{k}$ is the uncorrelated nonlinear distortion term, and
$\alpha_{k}$ is the complex attenuation factor that can be expressed
as

\begin{equation}
\ensuremath{\alpha_{k}=\beta_{1}+\beta_{3}T_{k}^{(3)}+\beta_{5}T_{k}^{(5)}}\label{eq:alpha_k}
\end{equation}
where 

\begin{equation}
\ensuremath{T_{k}^{(3)}=2\sum\limits _{l=0}^{N-1}\left|S_{l}\right|^{2}-\left|S_{k}\right|^{2}}
\end{equation}
and

\begin{align}
\ensuremath{T_{k}^{(5)}} & =6\sum\limits _{l=0}^{N-1}\sum\limits _{t=0}^{N-1}\left|S_{l}\right|^{2}\left|S_{t}\right|^{2}-6\sum\limits _{t=0}^{N-1}\left|S_{t}\right|^{2}\left|S_{k}\right|^{2}\nonumber \\
 & -3\sum\limits _{t=0}^{N-1}\left|S_{t}\right|^{4}+4\left|S_{k}\right|^{4}
\end{align}

Finally, the uncorrelated distortion term $d_{k}$ can be expressed
by 

\begin{equation}
\ensuremath{d_{k}=\beta_{3}\left[d_{k}^{(3)}-T_{k}^{(3)}S_{k}\right]+\beta_{5}\left[d_{k}^{(5)}-T_{k}^{(5)}S_{k}\right]}
\end{equation}
where $\ensuremath{d_{k}^{(3)}}$ and $\ensuremath{d_{k}^{(5)}}$
are calculated by (\ref{eq:Convolutions}). It worth noting that in
case of \emph{m}-QAM signaling $\alpha_{k}$ depends on all transmitted
symbols $\ensuremath{\{{S_{i}}\},{\rm {}}i=0,1,...,N-1}$. However,
it can be observed that for $N\rightarrow\infty$, $\alpha_{k}$ becomes
very close to its average value $\ensuremath{{\alpha_{k}}\approx\alpha}$.
If all symbols are transmitted independently and with equal probability,
we can obtain for very large \emph{N}

\begin{equation}
\ensuremath{\alpha\approx\beta_{1}+\beta_{3}T^{(3)}+\beta_{5}T^{(5)}}
\end{equation}
where

\begin{equation}
\ensuremath{T^{(3)}=\left(2N-1\right)E\left[\left|s_{l}\right|^{2}\right]},
\end{equation}

\begin{equation}
\ensuremath{T^{(5)}=6N(N-1)\left(E\left[\left|S_{l}\right|^{2}\right]\right)^{2}-(3N-4)E\left[\left|S_{l}\right|^{4}\right]},
\end{equation}
and $E[\cdot]$ denotes expectation. $\ensuremath{E\left[\left|S_{l}\right|^{2}\right]}$
and $\ensuremath{E\left[\left|S_{l}\right|^{4}\right]}$ constants
depend on constellation type.

In case of \emph{m}-QAM signaling with in-phase and quadrature components
$\ensuremath{I,Q=\left\{ -(\sqrt{m}-1),{\rm }-(\sqrt{m}-3),...{\rm },{\rm }+(\sqrt{m}-3),{\rm }+(\sqrt{m}-1)\right\} }$
straightforward calculation gives

\begin{equation}
\ensuremath{E\left[\left|S_{l}\right|^{2}\right]=\frac{2}{3}\left(m-1\right)}
\end{equation}

\begin{equation}
\ensuremath{E\left[\left|S_{l}\right|^{4}\right]=\frac{4}{45}\left(m-1\right)\left(7m-13\right)}
\end{equation}

\balance


\begin{thebibliography}{10}
\bibitem{key-43}Information Technology \textendash{} Telecommunications
And Information Exchange Between Systems \textendash{} Local and Metropolitan
Area Networks \textendash{} Specific Requirements \textendash{} Part
11: Wireless LAN Medium Access Control (MAC) and Physical Layer (PHY)
Specifications, IEEE Standard 802.11ac, 2013

\bibitem{key-42}J. Gozalvez, M. Sepulcre and R. Bauza, \textquotedbl{}IEEE
802.11p vehicle to infrastructure communications in urban environments,\textquotedbl{}
IEEE Communications Magazine, vol. 50, no. 5, pp. 176-183, May 2012

\bibitem{key-41}Frame structure channel coding and modulation for
a second generation digital terrestrial television broadcasting system
(DVB-T2), ETSI Standard 302 755, 2015

\bibitem{key-40}J. Andrews, S. Buzzi, W. Choi, S. Hanly, A. Lozano,
A. Soong, J. Zhang, ``What will 5G be?'', IEEE Journal On Selected
Areas in Commun., Vol. 32, No. 6, pp. 1065-1082, June 2014

\bibitem{key-33}P. Banelli and S. Cacopardi, \textquotedbl{}Theoretical
analysis and performance of OFDM signals in nonlinear AWGN channels\textquotedbl{},
IEEE Trans. Commun., Vol. 48, No. 3, pp. 430-441, Mar. 2000

\bibitem{key-34}D. Dardari, V. Tralli and A. Vaccari, \textquotedbl{}A
theoretical characterization of nonlinear distortion effects in OFDM
systems,\textquotedbl{} IEEE Trans. Commun, Vol. 48, No. 10, pp. 1755-1764,
Oct. 2000

\bibitem{key-35}P. Banelli, \textquotedbl{}Theoretical analysis and
performance of OFDM signals in nonlinear fading channels,\textquotedbl{}
IEEE Trans. Wireless Commun., Vol. 2, No. 2, pp. 284-293, Mar. 2003

\bibitem{key-58}J. H. Jong and W. E. Stark, \textquotedbl{}Performance
analysis of coded multicarrier spread-spectrum systems in the presence
of multipath fading and nonlinearities,\textquotedbl{} IEEE Trans.
Commun., Vol. 49, No. 1, pp. 168-179, Jan. 2001 

\bibitem{key-51}S.V. Zhidkov, ``Performance Analysis of Multicarrier
Systems in the Presence of Smooth Nonlinearity,'' EURASIP Journal
on Wireless Communications and Networking, Vol. 2, 2004, pp. 335\textendash 343

\bibitem{key-50}R. Enright and M. Darnell, \textquotedbl{}OFDM modem
with peak-to-mean power ratio reduction using adaptive clipping\textquotedbl{},
in Proc. IEE conf. HF radio systems and techniques, pp. 44-49, 1997 

\bibitem{key-20}T. A. Wilkinson and A. E. Jones, \textquotedbl{}Minimization
of the peak-to-mean envelope power ratio of multicarrier transmission
schemes by block coding\textquotedbl{}, in Proc. IEEE Vehicular Technology
Conf., Vol. 2, Chicago, IL, pp. 825-829, July 1995 

\bibitem{key-21}D. Han and T. Hwang, \textquotedbl{}An adaptive pre-distorter
for the compensation of HPA nonlinearity\textquotedbl{}, IEEE Trans.
Broadcasting, Vol. 46., pp 152-157, Jun. 2000 

\bibitem{key-25}J. Guerreiro, R. Dinis and P. Montezuma, \textquotedbl{}Optimum
and Sub-Optimum Receivers for OFDM Signals with Strong Nonlinear Distortion
Effects,\textquotedbl{} in IEEE Trans. on Commun., Vol. 61, No. 9,
pp. 3830-3840, Sept. 2013

\bibitem{key-12}J. Guerreiro, R. Dinis and P. Montezuma, \textquotedbl{}On
the Optimum Multicarrier Performance With Memoryless Nonlinearities,\textquotedbl{}
IEEE Trans. on Commun., Vol. 63, No. 2, pp. 498-509, Feb. 2015 

\bibitem{key-13}J. Tellado, L. M. C. Hoo and J. M. Cioffi, \textquotedbl{}Maximum-likelihood
detection of nonlinearly distorted multicarrier symbols by iterative
decoding\textquotedbl{}, IEEE Trans. Commun., Vol. 51, No. 2, pp.
218-228, Feb. 2003 

\bibitem{key-14}H. Chen and A. M. Haimovich, \textquotedbl{}Iterative
estimation and cancellation of clipping noise for OFDM signals,\textquotedbl{}
IEEE Commun. Lett., Vol. 7, No. 7, pp. 305-307, July, 2003 

\bibitem{key-15}D. Kim and G. L. Stüber, \textquotedbl{}Clipping
noise mitigation for OFDM by decision-aided reconstruction\textquotedbl{},
IEEE Commun. Lett., Vol. 3, No. 1, pp. 4-6, Jan. 1999

\bibitem{key-17}L. G. Baltar, S. Dierks, F. H. Gregorio, J. E. Cousseau
and J. A. Nossek, \textquotedbl{}OFDM receivers with iterative nonlinear
distortion cancellation,\textquotedbl{} in Proc. IEEE Eleventh International
Workshop on Signal Processing Advances in Wireless Communications
(SPAWC), Marrakech, 2010, pp. 1-5

\bibitem{key-18}F. H. Gregorio, S. Werner, J. Cousseau, J. Figueroa,
R. Wichman, ``Receiver-side nonlinearities mitigation using an extended
iterative decision-based technique,'' Signal Processing, Vol. 91,
Issue 8, pp. 2042-2056, August 2011

\bibitem{key-19}C. Rapp, \textquotedbl{}Effects of HPA-nonlinearity
on a 4-DPSK/ OFDM-signal for a digital sound broadcasting system\textquotedbl{},
Proc. of the Second European Conf. on Satellite Comm., Belgium, pp.
179-184, Oct. 1991 

\bibitem{key-20}A. Saleh, \textquotedbl{}Frequency-independent and
frequency-dependent nonlinear models of TWT amplifiers\textquotedbl{},
IEEE Trans. Commun., vol. 29, pp. 1715-1720, Nov. 1981 

\bibitem{key-21}A. Ghorbani, and M. Sheikhan, \textquotedbl{}The
effect of Solid State Power Amplifiers (SSPAs) Nonlinearities on MPSK
and M-QAM Signal Transmission\textquotedbl{}, Sixth Int'l Conference
on Digital Processing of Signals in Comm., 1991, pp. 193-197

\bibitem{key-30}K. G. Gard, H. M. Gutierrez and M. B. Steer, \textquotedbl{}Characterization
of spectral regrowth in microwave amplifiers based on the nonlinear
transformation of a complex Gaussian process,\textquotedbl{} IEEE
Trans. Microwave Theory and Techniques, Vol. 47, No. 7, pp 1059-1069,
July 1999 

\bibitem{key-31}H. Gutierrez, K. Gard and M. B. Steer, \textquotedbl{}Nonlinear
gain compression in microwave amplifiers using generalized power-series
analysis and transformation of input statistics,\textquotedbl{} IEEE
Trans. on Microwave Theory and Techniques, Vol. 48, No. 10, pp. 1774-1777,
Oct. 2000 

\bibitem{key-32}G. T. Zhou and J. S. Kenney, \textquotedbl{}Predicting
Spectral Regrowth of Nonlinear Power Amplifiers,\textquotedbl{} IEEE
Trans. on Commun., Vol. 50, No. 5, pp. 718-722, May 2002 

\bibitem{key-39}N. M. Blachman, \textquotedbl{}The Output Signals
and Noise from a Nonlinearity with Amplitude-Dependent Phase Shift,\textquotedbl{}
IEEE Trans. on Information Theory, Vol. 25, No. 1, pp. 77-79, Jan.
1979 

\bibitem{key-36}A. V. Oppenheim, R. W. Schaffer and J. R. Buck, Discrete-time
signal processing, 2nd ed., New Jersey: Prentice-Hall Inc., 1998 

\bibitem{key-55}S.M. Kay, Fundamentals of statistical signal processing:
estimation theory, New Jersey: Prentice-Hall Inc., 1993

\bibitem{key-59}J. Medbo and P. Schramm, \textquotedblleft Channel
models for HIPERLAN/2 in different indoor scenarios,\textquotedblright{}
ETSI Broadband Radio Access Networks, Document no. 3ERI085B, Mar.1998 

\bibitem{key-5}P. Gilabert, G. Montoro and E. Bertran, \textquotedbl{}On
the Wiener and Hammerstein models for power amplifier predistortion,\textquotedbl{}
in Asia-Pacific Microwave Conference Proceedings, 2005, pp. 4-6

\bibitem{key-38}H. Stark and J. W. Woods, Probability, random processes,
and estimation theory for engineers, 2nd ed., New Jersey: Prentice-Hall
Inc., 1994 
\end{thebibliography}
\end{document}